\documentclass[pre, twocolumn, showpacs, superscriptaddress]{revtex4}
\usepackage{amsmath}
\usepackage{dcolumn}
\usepackage[xdvi]{graphicx}%
\makeatletter

\def\btt#1{\texttt{\@backslashchar#1}}%
\DeclareRobustCommand\bblash{\btt{\@backslashchar}}%
\makeatother

\newcommand{\bv}[1]{{\boldsymbol #1}}

\begin{document}

\title{Critical behaviors of sheared frictionless granular materials 
near jamming transition }

\author{ Michio Otsuki$^1$ and Hisao Hayakawa$^2$}
\affiliation{
$^1$ Department of Physics and Mathematics, Aoyama Gakuin University,
5-10-1 Fuchinobe, Sagamihara, Kanagawa, 229-8558, Japan. \\
  $^2$ Yukawa Institute for Theoretical Physics, Kyoto University,  Kitashirakawaoiwake-cho, Sakyo-ku, Kyoto 606-8502, Japan.}

\begin{abstract}
Critical behaviors of sheared dense and frictionless granular materials in the vicinity of  the jamming transition are numerically investigated.
From the extensive molecular dynamics simulation,
we verify the validity of the scaling theory near the jamming transition 
proposed by Otsuki and Hayakawa
(Prog. Theor. Phys., {\bf 121}, 647 (2009)).
We also clarify the critical behaviors of  the shear viscosity
 and the pair correlation function based on both a mean field theory and the simulation.
\end{abstract}

\pacs{45.70.-n, 05.70.Jk, 47.50.-d}

\maketitle

\section{Introduction}

Let us consider mechanical properties of grains which are packed 
in a container. 
When the density is low enough
and the effect of gravity is negligible, 
there is no pressure acting on the wall of the container.
However, when the density exceeds a critical value, the pressure acting on the wall becomes finite.
Such kind of transition for the rigidity or the stress is known as the jamming transition.

Jamming is a key concept to characterize the transition of athermal systems such as
granular materials \cite{Jaeger},
colloidal suspensions \cite{Pusey}, emulsions and forms \cite{Durian}.
Liu and Nagel 
suggested that  a unifying description might be possible to cover both glass transition in thermal systems and
athermal jamming transition \cite{Liu}.
Indeed, there are many similarities between the jamming 
and the glass transition.
For example, 
the rheological properties of the jammed systems \cite{Hatano07,Olsson}
are similar to those of glassy materials \cite{Berthier02}, and
the granular materials near the jamming transition point 
(point J) exhibit the dynamical heterogeneity which is one of the 
characteristics of glassy materials
\cite{Abate06, Abate07, Dauchot08, Watanabe, Olsson, Hatano08.2, Hatano08.3}.
From these similarities, the mode coupling theory for glassy materials
\cite{Fuchs, Miyazaki, Chong}  is applied to granular materials,
but it fails to describe the jamming transition \cite{Hayakawa}.


Recent studies have revealed that the jamming is the transition to appear in
the pressure, the coordination number, the elastic moduli \cite{OHern02,OHern03},
and the soft modes \cite{Wyart}.
In particular, Olsson and Teitel \cite{Olsson}, and Hatano \cite{Hatano08} numerically 
found scaling relations
for the shear stress, the pressure, and the kinetic temperature,
  which are characterized by some critical exponents.
A similar scaling relation is observed in the simulation 
of Josephson junction arrays \cite{Yoshino09}.
These scaling relations indicate that jamming is a continuous transition in which the stress is zero at the critical point.
In order to understand the critical behaviors of the jamming transition, 
it is important to determine the critical exponents.
In the previous paper \cite{Otsuki08}, the authors theoretically obtained 
the critical exponents for the scaling relations 
of the 
shear stress, the kinetic temperature and the characteristic frequency,
which characterizes the collisional energy loss,
for sheared frictionless granular materials. They also 
numerically verified the validity of their theoretical predictions
for the linear spring model \cite{Otsuki08}.


In this paper, we will verify the validity of the predictions 
in Ref. \cite{Otsuki08}
for sheared frictionless granular materials
  in various situations. We will also discuss the behaviors of the viscosity
  and the pair correlation function in details. 
In the next section, we will summarize the theoretical predictions by the present authors \cite{Otsuki08}.
In Sec. \ref{Simulation:sec}, we will compare the results of our extensive simulations with the theoretical predictions.
Section \ref{Simulation:sec} consists of five subsections:
Section \ref{setup:subsec} will be devoted to 
the explanation of the setup of our simulations, in Sec.\ref{scaling:subsec} 
we will present various scaling plots to verify
the scaling theory,  in Sec. \ref{delta:subsec} we will discuss 
the force law dependence of the critical exponents, 
we will explain the density dependence of 
critical variables in Sec. \ref{density:subsec}. 
We will explain the results for nearly elastic cases 
in Sec. \ref{elastic:subsec}.  
 Section \ref{pair-correlation} will be devoted to the explanation of
 critical properties of the pair correlation function in the vicinity of the jamming transition,
which was not discussed in Ref. \cite{Otsuki08}.
In Sec. \ref{Discussion:sec}, we will discuss and conclude our results.
In Appendix A, we will summarize the method for the theoretical prediction which contains some generalized results beyond Ref. \cite{Otsuki08} based on a
different method for the derivation.
In Appendix B, we will derive some relations, which are necessary to
discuss the critical behavior of the pair correlation function.

\section{Mean field theory}
\label{Theory:sec}

In this section, we briefly summarize the theoretical results
in Ref. \cite{Otsuki08}.
Let us consider $D$-dimensional granular assemblies 
under an uniform shear with shear rate $\dot \gamma$.
The system includes $N$ grains, each of which has the identical mass $m$.
The packing fraction of the system and the critical fraction at point J are respectively denoted by  $\phi$ and  $\phi_J$.
Throughout this paper, we assume that granular particles are frictionless, 
where any contact force acts along the line to connect two centers of mass of contacting grains. 
In most of our arguments,  we assume that the elastic interaction between the grain $i$ located at $\bv{r}_i$ and the grain $j$ at $\bv{r}_j$
is given by
\begin{eqnarray}
f_{\rm el}(r_{ij}) & = & k \Theta \left( \sigma_{ij} - r_{ij} \right )(\sigma_{ij}-r_{ij})^\Delta,
\label{elastic:force}
\end{eqnarray}
where $k$ and $r_{ij}$ are the elastic constant and the distance between 
the grains $r_{ij}\equiv |\bv{r}_{ij}|=|\bv{r}_i-\bv{r}_j|$, respectively.
$\sigma_{ij} = (\sigma_i + \sigma_j)/2$ is the average of the diameter
of the grain $i$ and the grain $j$ with diameters $\sigma_i$ and $\sigma_j$.
$\Theta(x)$ is the Heaviside step function satisfying $\Theta(x) = 1$
for $x \geq 0$ and $\Theta(x) = 0$ for otherwise.
 Many researchers use 
the linear spring model ($\Delta=1$) associated with the viscous contact force 
\begin{eqnarray}
f_{\rm vis}(r_{ij}, v_{ij,{\rm n}}) & = & 
- \eta \Theta \left( \sigma_{ij} - r_{ij} \right ) 
v_{ij,{\rm n}},
\label{dis:lin}
\end{eqnarray}
where $\eta$ is the viscous parameter. Here, $v_{ij,{\rm n}}$ is
the relative normal velocities between the grains given by 
$v_{ij,{\rm n}}\equiv (\bv{v}_i-\bv{v}_j)\cdot\bv{r}_{ij}/r_{ij}$,
where $\bv{v}_i$ and $\bv{v}_j$ are the velocities of the grain $i$
and the grain $j$, respectively.
On the other hand, the Herzian model (Eq.(\ref{elastic:force}) with $\Delta=3/2$)  associated with
the corresponding viscous contact force
\begin{eqnarray}
f_{\rm vis}(r_{ij}, v_{ij,\parallel}) & = & 
- \eta \Theta \left( \sigma_{ij} - r_{ij} \right ) 
 \sqrt{\sigma_{ij}-r_{ij}}
v_{ij,{\rm n}}
\label{dis:her}
\end{eqnarray}
is more realistic one for 
three-dimensional spheres.


In the previous paper \cite{Otsuki08}, we introduced the scaling relations
\begin{eqnarray}
T & = & A_{T,D}|\Phi|^{x_{\Phi}} {\cal T}_{\pm}\left(t_D\frac{\dot\gamma}{|\Phi|^{x_{\Phi}/x_{\gamma}}}\right), \label{T:scale} \\
S & = & A_{S,D}|\Phi|^{y_{\Phi}}{\cal S}_{\pm}\left(s_D\frac{\dot\gamma}{|\Phi|^{y_{\Phi}/y_{\gamma}}}\right), \label{S:scale}\\
P & = & A_{P,D}|\Phi|^{y_{\Phi}'}{\cal P}_{\pm}\left(p_D\frac{\dot\gamma}{|\Phi|^{y_{\Phi}'/y_{\gamma}'}}\right), 
\label{P:scale} \\
\omega & = & A_{W,D}|\Phi|^{y_{\Phi}}{\cal W}_{\pm}\left(w_D\frac{\dot\gamma}{|\Phi|^{y_{\Phi}/y_{\gamma}}}\right), 
\label{w:scale} 
\end{eqnarray}
where $\Phi \equiv  \phi - \phi_J$. The kinetic temperature $T$, 
the shear stress $S$, and the pressure $P$ are respectively given by 
\cite{Evans}
\begin{eqnarray}
T & = & \frac{1}{ND}\left <   \sum_{i=1}^N \frac{|\bv{p}_i|^2}{2m}\right >,  
\label{T:calc}\\
S & = &  -\frac{1}{V}\left <    \sum_i^N \sum_{j>i} \frac{r_{ij,x} r_{ij,y}}{r_{ij}}
 \left [ f_{{\rm el}}(r_{ij}) + 
f_{{\rm vis}}(r_{ij}, v_{ij,\parallel}) \right ]
\right > 
\nonumber \\
& &
-\frac{1}{V} \left <   \sum_{i=1}^N \frac{p_{x,i}p_{y,i}}{2m} \right >
\label{S:calc},  \\
P & = & 
\frac{1}{DV} \left < \sum_i^N \sum_{j>i} r_{ij}
 \left [ f_{{\rm el}}(r_{ij}) + 
f_{{\rm vis}}(r_{ij}, v_{ij,\parallel}) \right ]
\right >  \nonumber \\
& & +
\frac{1}{DV} \left <   \sum_{i=1}^N \frac{|\bv{p}_i|^2}{2m} \right >,
\label{P:ex}
\end{eqnarray}
where we have introduced
$\bv{p}_i \equiv m [ \bv{v}_i - \bv{c}(\bv{r}_i)]$ 
with the average velocity $c_\alpha(\bv{r})= \dot \gamma y \delta_{\alpha, x}$ 
at the position $\bv{r}$, 
$V$ is the volume of the system, and
$\left < \cdot \right >$ represents the ensemble average.
The characteristic frequency $\omega$ is defined by
\begin{eqnarray}\label{omega}
\omega \equiv \frac{\dot{\gamma}S}{nT},
\end{eqnarray}
where $n$ is the number density.
This $\omega$ is reduced to the collision frequency in the unjammed phase determined by the balance between
the viscous heating and the collisional energy loss. 
${\cal T}_{+}(x)$, ${\cal S}_{+}(x)$, ${\cal P}_{+}(x)$ and ${\cal W}_{+}(x)$
are the scaling functions above point J (in the jammed phase).
On the other hand, 
the scaling functions below point J (in the unjammed phase) are given by
${\cal T}_{-}(x)$, ${\cal S}_{-}(x)$, ${\cal P}_{-}(x)$ and ${\cal W}_{-}(x)$.
Equations (\ref{T:scale})--(\ref{w:scale}) contain the critical exponents
$x_\phi$, $x_\gamma$, $y_\phi$, $y_\gamma$, $y_\phi'$, $y_\gamma'$, 
$z_\phi$, and $z_\gamma$.
It should be noted that 
$A_{T,D}$, $A_{S,D}$, $A_{W,D}$, $A_{P,D}$, $t_D$, $s_D$, $w_D$, and $p_D$
are the constants depending only on the dimension $D$.

From the scaling theory explained in Appendix A,
there are the scaling relations in the unjammed phase 
in the limit $\dot \gamma \rightarrow 0$ as
\begin{eqnarray} \label{unjammed:scaling}
T & \sim & \dot \gamma^2 |\Phi|^{x_\phi ( 1- 2/x_\gamma)}, \quad
S \sim \dot \gamma^2 |\Phi|^{y_\phi ( 1- 2/y_\gamma)}, \nonumber \\
P & \sim & \dot \gamma^2 |\Phi|^{y_\phi' ( 1- 2/y_\gamma')}, \quad
\omega \sim \dot \gamma |\Phi|^{z_\phi ( 1- 1/z_\gamma)}. 
\end{eqnarray}
Similarly, $T$, $S$, $P$, and $\omega$ satisfy
\begin{eqnarray} \label{jammed:scaling}
T & \sim & \dot \gamma |\Phi|^{x_\phi ( 1- 1/x_\gamma)}, \quad
S \sim |\Phi|^{y_\phi }, \nonumber \\
P & \sim & |\Phi|^{y_\phi' }, \quad
\omega \sim |\Phi|^{z_\phi },
\end{eqnarray}
in the zero shear limit of the jammed phase.
The scaling relations at point J, i.e. $\Phi \simeq 0$,  are given by
\begin{equation} \label{critical:scaling}
T \sim \dot \gamma^ {x_\gamma}, \quad
S \sim \dot \gamma^ {y_\gamma}, \quad
P \sim \dot \gamma^ {y_\gamma'}, \quad
\omega \sim \dot \gamma^ {z_\gamma}.
\end{equation}

As explained in Appendix A and Ref. \cite{Otsuki08}, 
the critical exponents are given by
\begin{eqnarray}\label{exponents}
x_{\Phi}&=&2+\Delta, \quad x_{\gamma}=\frac{2\Delta+4}{\Delta+4},
 \quad y_{\Phi}=\Delta, \quad y_{\gamma}=\frac{2\Delta}{\Delta+4},
\nonumber\\
 y_{\Phi}'& =& \Delta, \quad y_{\gamma}'=\frac{2\Delta}{\Delta+4}, \quad
z_{\Phi}=\frac{\Delta}{2}, \quad 
z_{\gamma}=\frac{\Delta}{\Delta+4}.
\end{eqnarray} 
Note that the scaling of the pressure and the determination of the exponents  $y_\gamma'$ 
were not discussed in Ref. \cite{Otsuki08}.  The derivations which are a little different from
the original one are explained in Appendix A.
We should stress that the exponents are independent of the spatial dimension $D$,
and the characteristic feature of the jamming transition is
the $\Delta$-dependence of the critical exponents.
Indeed, O'Hern et al. found that the pressure behaves as $P \sim |\Phi|^\Delta$
for unsheared jammed systems in the vicinity of the jamming point.
Hatano also indicated that the critical exponents for the sheared granular materials
differ in the cases of $\Delta = 1$ and $3/2$.
Explicit $\Delta$-dependence 
of the exponents in Eq. (\ref{exponents}) except for $y_{\gamma}'$ for the sheared granular materials was obtained 
in Ref. \cite{Otsuki08}.
It should be noted that  
our theory can be generalized to the case of the contact force
\begin{equation}
f_{\rm el}(r_{ij})=\Theta(\sigma_{ij}-r_{ij}){\cal F}(r_{ij}).
\label{general:force}
\end{equation}
Here, the exponent $\Delta$ in Eq. \eqref{exponents} should be determined from the relation
$\lim_{r_{ij} \rightarrow \sigma_{ij}} {\cal F}(r_{ij})
\propto (\sigma_{ij} - r_{ij})^\Delta$ for this case.
If we use an analytic function ${\cal F}(r_{ij})$ 
such as the repulsive Lennard-Jones force
\begin{equation}
{\cal F}(r_{ij}) = \frac{12\epsilon}{\sigma_{ij}} \left [ 
\left ( \frac{\sigma_{ij}}{r_{ij}} \right )^{13}
-\left ( \frac{\sigma_{ij}}{r_{ij}} \right )^{7}
\right ], \label{repLJ}
\end{equation}
the value of $\Delta$ should be $\Delta=1$, because we can use the expansion
${\cal F}(r_{ij})\simeq {\cal F}'(\sigma_{ij})(\sigma_{ij}-r_{ij})+{\cal F}"(\sigma_{ij})/2(\sigma_{ij}-r_{ij})^2+\cdots$.
Note that Eq. (\ref{repLJ}) differs from the usual form
for the Lennard-Jones potential in order to satisfy ${\cal F}(\sigma_{ij}) = 0$.

\section{Comparison between theory and simulation}
\label{Simulation:sec}

In this section, let us verify the validity of the theory \cite{Otsuki08}
by the molecular dynamics simulation.
As stated in Introduction,  we have already confirmed that
all of our results of the simulation for $\Delta=1$ under one size distribution of grains with fixing the strength of inelasticity
are consistent with the theory \cite{Otsuki08},
but nobody has checked its validity for general $\Delta$ under various situations.
Thus, we will perform extensive simulations with changing $\Delta$, the size distribution, and the inelasticity etc.
 
In the first part (\ref{setup:subsec}), we explain the details of our model.
In the second part (\ref{scaling:subsec}), we demonstrate the validity of the scaling theory
given by Eqs. \eqref{T:scale}--\eqref{w:scale} with Eq. \eqref{exponents} 
from the test of scaling plots under various conditions.
In the third part (\ref{delta:subsec}), we check  the theory 
for the $\Delta$-dependence of the critical exponents.
In the fourth part (\ref{density:subsec}),  we discuss  $\Phi$-dependences of the characteristic frequency
$\omega$ in the jammed  region and the viscosity $\mu= S/\dot \gamma$ in the unjammed region.
In the last part (\ref{elastic:subsec}), 
we explain the behaviors for nearly elastic cases, 
while the results in the first five parts are obtained for strongly dissipative cases.

\subsection{Setup}
\label{setup:subsec}

We examine three different systems on dispersion of diameters of grains.
The first system we call the {\it polydisperse} system 
consists of four types of grains, and the diameters of grains are
$0.7\sigma_0$, $0.8\sigma_0$, $0.9\sigma_0$ and $\sigma_0$, where 
the number of each type of grains is $N/4$.
The {\it polydisperse} system has been studied in Ref. \cite{Otsuki08},
where the critical fraction is estimated as 
$\phi_J = 0.84285$ for $D=2$, $\phi_J = 0.64455$ for $D=3$
or $\phi_J = 0.4615$ for $D=4$ \cite{Otsuki08}. 
The second system which we call the {\it bidisperse} system 
consists of  two types of grains. The diameters of grains 
are $5 \sigma_0 / 7$ and $\sigma_0$, where
the number of each type of grains is $N/2$.
The {\it bidisperse} system has been studied by many researchers 
\cite{OHern02,OHern03,Berthier,Inagaki},
and the critical density $\phi_J$ is known as $0.648$
for $D=3$ by the numerical simulation of static granular packings
  \cite{OHern02}.
The final system we call the {\it monodisperse} system consists of only one type of particles,
whose diameters are $\sigma_0$.
The critical density $\phi_J$ of the {\it monodisperse} system is believed to be $0.639$
for $D=3$ from a numerical simulation \cite{OHern03}.

The time evolution equations of the position $\bv{r}_i$ and 
the velocity $\bv{v}_i$ of the $i$th particle are given by
\begin{eqnarray} 
\frac{d \bv{r}_i}{dt} & = & \bv{v}_i, \\
m \frac{d \bv{v}_i}{dt} & = & \sum_{j\neq i} 
\left \{ f_{\rm el}(r_{ij}) + 
f_{\rm vis}(r_{ij}, v_{ij,{\rm n}}) \right \} \frac{\bv{r}_{ij}}{r_{ij}},
\end{eqnarray}
where 
the elastic force $f_{\rm el}(r_{ij})$ is given by Eq. 
(\ref{elastic:force}) for most of cases except for the 
case of the repulsive Lennard-Jones potential.
$f_{\rm vis}(r_{ij}, v_{ij,{\rm n}})$ 
given by Eqs. (\ref{dis:lin}) or (\ref{dis:her}) is the viscous contact force between particles.
In order to realize an uniform  
velocity gradient $\dot\gamma$ in  
$y$ direction and macroscopic velocity only 
in the $x$ direction, we adopt
the Lees-Edwards boundary conditions.

In our simulation with the elastic 
contact force given by Eq. \eqref{elastic:force}, 
$m$, $\sigma_0$ and $k$ are set to be unity, 
and all quantities are converted to dimensionless forms,
where the unit of time scale is $\sqrt{m \sigma_0^{1-\Delta} / k}$.
In the case of the contact force given by Eqs. \eqref{general:force}
and \eqref{repLJ},
$\epsilon$ in Eq. (\ref{repLJ}) is set to be unity instead of $k$,
and  the unit of time scale is $\sqrt{m \sigma_0^2 / \epsilon}$.
We adopt the elastic constant $k=1.0$ or $\epsilon=1.0$, and the viscous constant $\eta   = 1.0 $ for most of cases except for nearly elastic cases in Sec. \ref{elastic:subsec}.
This situation corresponds to the constant restitution coefficient $e=0.043$ for the linear spring model.
We use the leap-frog algorithm, which is second-order accurate in time, 
by using the time interval $\Delta t=0.2$ for the cases of
the contact force given by Eq. \eqref{elastic:force} 
with checking the convergence until $\Delta t=0.05$.  
In the simulation with the contact force given by Eqs. \eqref{general:force}
and \eqref{repLJ}, we use $\Delta t = 0.01$.

\subsection{Scaling plots}
\label{scaling:subsec}

\begin{figure}
\begin{center}
\includegraphics[height=12em]{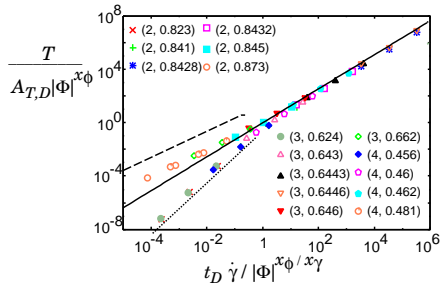}
\caption{ (Color online)
 Collapsed data of the shear rate dependence of the kinetic temperature $T$
 in the {\it polydisperse} system ($N=2000$)
 with $\Delta=3/2$ using the scaling law, Eq. (\ref{T:scale}), for $D=2,3$ and $4$.
The dashed line, the dotted line and the solid line are proportional to 
$\dot{\gamma}$, $\dot{\gamma}^2$ and $\dot{\gamma}^{x_\gamma}$, respectively.
 The legends show the dimension $D$ and the volume fraction $\phi$
 as $(D, \phi)$.
 The critical exponents  estimated from Eq. \eqref{exponents}
 are $x_\phi=9/2$ and $x_\gamma=14/11$.
  }
\label{T_H}
\end{center}
\end{figure}

\begin{figure}
\begin{center}
\includegraphics[height=12em]{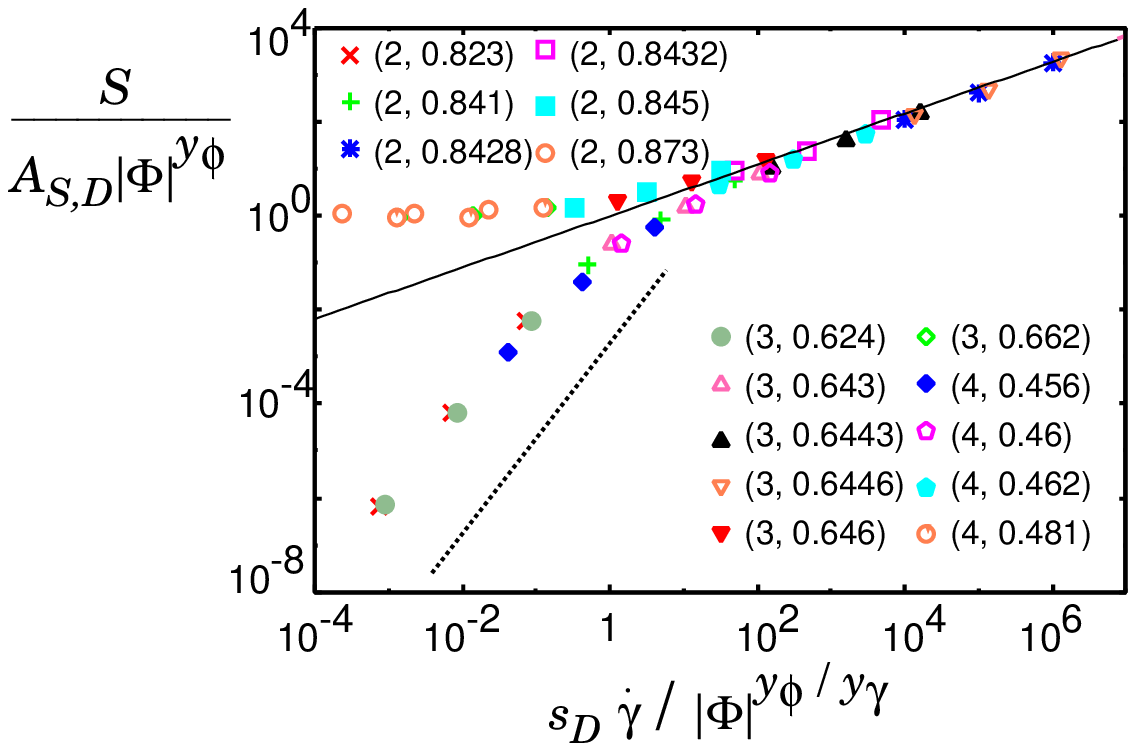}
\caption{ (Color online)
Collapsed data of the shear rate dependence of the shear stress $S$ 
 for the {\it polydisperse} system ($N=2000$) 
  with $\Delta=3/2$ using the scaling law, Eq. (\ref{S:scale}), 
  for $D=2,3$ and $4$.
The dotted line and the solid line are proportional to 
 $\dot{\gamma}^2$ and $\dot{\gamma}^{y_\gamma}$, respectively.
 The legends show the dimension $D$ and the volume fraction $\phi$
 as $(D, \phi)$.
 The critical exponents  estimated from Eq. \eqref{exponents}
 are $y_\phi=3/2$ and $y_\gamma=6/11$.
  }
\label{S_H}
\end{center}
\end{figure}

\begin{figure}
\begin{center}
\includegraphics[height=12em]{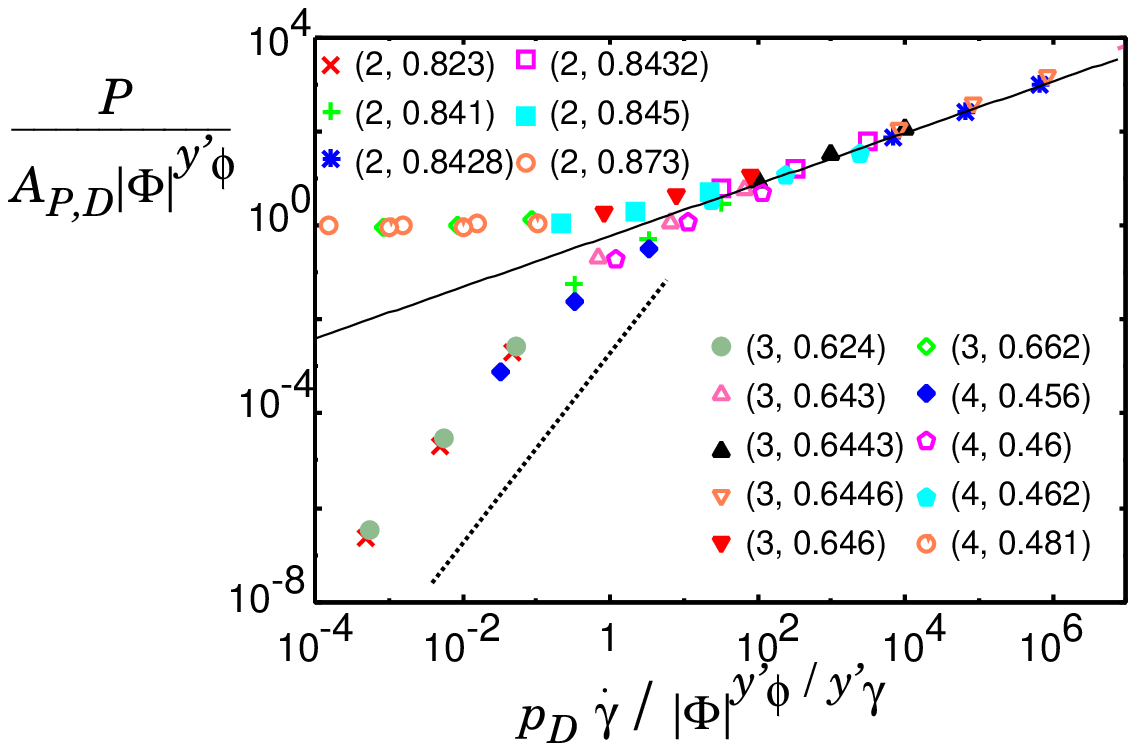}
\caption{ (Color online)
Collapsed data of the shear rate dependence of the pressure $P$
 for the {\it polydisperse} system ($N=2000$)
  with $\Delta=3/2$ using the scaling law,  Eq. (\ref{P:scale}),
   for $D=2,3$ and $4$.
The dotted line and the solid line are proportional to 
 $\dot{\gamma}^2$ and $\dot{\gamma}^{y_\gamma'}$, respectively.
 The legends show the dimension $D$ and the volume fraction $\phi$
 as $(D, \phi)$.
 The critical exponents estimated from Eq. \eqref{exponents}
 are $y'_\phi=3/2$ and $y'_\gamma=6/11$.
  }
\label{P_H}
\end{center}
\end{figure}

\begin{figure}
\begin{center}
\includegraphics[height=12em]{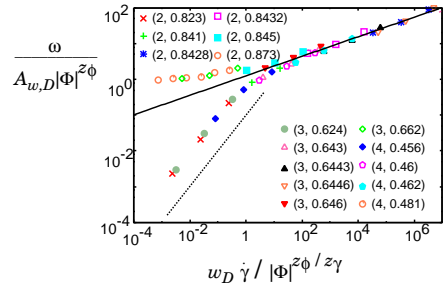}
\caption{ (Color online)
Collapsed data of the shear rate dependence of the characteristic frequency 
$\omega$ for the {\it polydisperse} system ($N=2000$)
  with $\Delta=3/2$ using the scaling law, Eq. (\ref{w:scale}),
   for $D=2,3$ and $4$.
The dotted line and the solid line are proportional to 
 $\dot{\gamma}$ and $\dot{\gamma}^{z_\gamma}$, respectively.
 The legends show the dimension $D$ and the volume fraction $\phi$
 as $(D, \phi)$.
 The critical exponents estimated from Eq. \eqref{exponents}
 are $z_\phi=3/4$ and $z_\gamma=3/11$.
  }
\label{w_H}
\end{center}
\end{figure}

Figures \ref{T_H}-\ref{w_H} show the scaling plots of
the {\it polydisperse} system 
with $\Delta=3/2$ 
based on Eqs. \eqref{T:scale}--\eqref{w:scale} with Eq. \eqref{exponents}
for the various dimensions $D=2,3$, and $4$.
(See Ref. \cite{Otsuki08} for the scaling plots of the 
{\it polydisperse} system with $\Delta=1$).
We use Eq. \eqref{dis:her}  for the viscous contact force.
Here, the number of the particles $N$ is $2000$,
and the shear rate $\dot{\gamma}$ is ranged between
$5\times 10^{-7}$ and $5 \times 10^{-5}$ for $D=2,3$ and between $5\times 10^{-6}$ and $5 \times 10^{-4}$ for $D=4$.
We also use the amplitudes and the adjustable parameters
$(t_D,A_{t,D},s_D,A_{s,D},p_D,A_{p,D},w_D,A_{w,D}) = 
(0.01, 15.0, 0.03, 0.02, 0.02, 0.3, 0.1, 0.12)$ for $D=2$,   
$(0.01, 6.0, 0.04, 0.03, 0.025, 0.3, 0.15, 0.25)$ for $D=3$,   
$(0.1, 0.45, 0.05, 0.05, 0.04, 0.5, 0.1, 0.45)$ for $D=4$.
All of the data converge to the universal master curves.
These results verify the validity of our theoretical predictions in Eq. \eqref{exponents}.

\begin{figure}
\begin{center}
\includegraphics[height=12em]{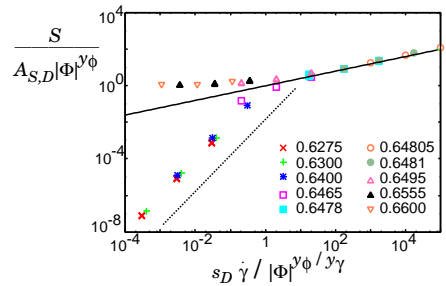}
\caption{ (Color online)
 Collapsed data of the shear rate dependence of the shear stress $S$
 for the {\it bidisperse} system ($N=4000$)
with  $\Delta=1$ using the scaling law, Eq. (\ref{S:scale}),
 for $D=3$.
The dotted line and the solid line are proportional to 
$\dot{\gamma}^2$ and $\dot{\gamma}^{y_\gamma}$, respectively.
 The legends represent the volume fraction $\phi$.
 The critical exponents estimated from Eq. \eqref{exponents}
 are $y_\phi=1$ and $y_\gamma=2/5$.
  }
\label{S_L_2dis}
\end{center}
\end{figure}

\begin{figure}
\begin{center}
\includegraphics[height=12em]{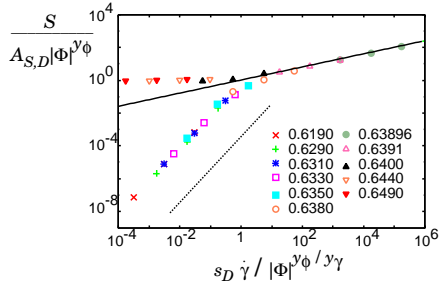}
\caption{ (Color online)
 Collapsed data of the shear rate dependence of the shear stress $S$
 for the {\it monodisperse} system ($N=4000$)
with $\Delta=1$ using the scaling law, Eq.  (\ref{S:scale}), 
for $D=3$.
The dotted line and the solid line are proportional to 
$\dot{\gamma}^2$ and $\dot{\gamma}^{y_\gamma}$, respectively.
 The legends show the volume fraction $\phi$.
 The critical exponents estimated from Eq. \eqref{exponents}
 are $y_\phi=1$ and $y_\gamma=2/5$.
  }
\label{S_mono}
\end{center}
\end{figure}

In Figs. \ref{S_L_2dis} and \ref{S_mono},
we show the scaling plots of $S$
for the {\it bidisperse} system and the {\it monodisperse} system with $\Delta=1$
and $D=3$, respectively, where both systems include 
 $N=4000$ particles.
The viscous contact force is given by Eq. \eqref{dis:lin},
and the shear rate $\dot{\gamma}$ is ranged between
$5\times 10^{-7}$ and $5 \times 10^{-5}$. 
The amplitude and the adjustable parameter are
$(s_D,A_{s,D}) = 
(0.035, 0.04)$ for  the {\it bidisperse} system
and $(0.025, 0.035)$ for  the {\it monodisperse} system.
These  scaling plots support the validity of our prediction.
The scaling plots for $T$, $P$, and $\omega$
for the {\it bidisperse} and the {\it monodisperse} systems also exhibit  
  elegant scalings, 
but we omit these figures in this paper.

\begin{figure}
\begin{center}
\includegraphics[height=12em]{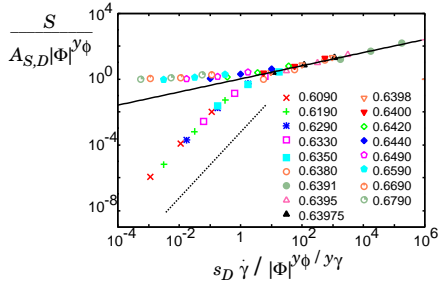}
\caption{ (Color online)
 Collapsed data of the shear rate dependence of the shear stress $S$
 in the larger {\it monodisperse} system ($N=20000$)
with $\Delta=1$ using the scaling law, Eq. (\ref{S:scale}),
 for $D=3$.
The dotted line and the solid line are proportional to 
$\dot{\gamma}^2$ and $\dot{\gamma}^{y_\gamma}$, respectively.
 The legends show the volume fraction $\phi$.
 The critical exponents estimated from Eq. \eqref{exponents}
 are $y_\phi=1$ and $y_\gamma=2/5$.
  }
\label{S_Lsize}
\end{center}
\end{figure}

We have checked the validity of our scaling theory 
in larger systems. Figure \ref{S_Lsize} shows the scaling plot
of $S$ in the three-dimensional {\it monodisperse} system
with $N=20000$ particles, where the shear rate $\dot{\gamma}$ is ranged between
$5\times 10^{-6}$ and $5 \times 10^{-4}$. 
The parameters and the guide lines are the same as
those for Fig. \ref{S_mono}. This scaling  supports 
the validity of our theory even in the larger system.

\begin{figure}
\begin{center}
\includegraphics[height=12em]{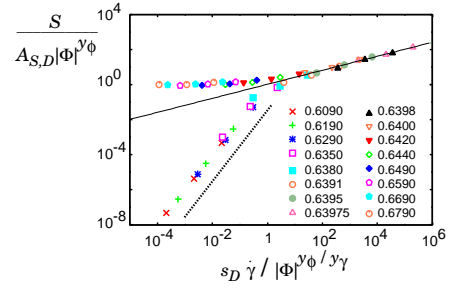}
\caption{ (Color online)
 Collapsed data of the shear rate dependence of the shear stress $S$
 in the {\it monodisperse} system ($N=2000$)
with the contact force given by Eqs. \eqref{general:force} and \eqref{repLJ} 
using the scaling law, Eq. \eqref{S:scale} for $D=3$.
The dotted line and the solid line are proportional to 
$\dot{\gamma}^2$ and $\dot{\gamma}^{y_\gamma}$, respectively.
 The legends show the volume fraction $\phi$.
 The critical exponents estimated from Eq. \eqref{exponents}
 with $\Delta=1$
 are $y_\phi=1$ and $y_\gamma=2/5$.
  }
\label{S_LJ}
\end{center}
\end{figure}

In order to verify the validity of our theory
in more general cases than that of Eq. (\ref{elastic:force}),
we examine the scaling plot for $S$ in the three-dimensional {\it monodisperse} 
system with the elastic contact force given by Eqs. \eqref{general:force} and 
\eqref{repLJ} 
and the viscous contact force given by Eq. \eqref{dis:lin}
in Fig. \ref{S_LJ}. Here we use
the number of the particles $N=2000$, and the shear rate $\dot \gamma$
is ranged between $10^{-5}$ and $10^{-3}$. 
The amplitude and the adjustable parameter are given by
$(s_D,A_{s,D}) = 
(0.004, 3.0)$, where we adopt the exponents given by Eq. \eqref{exponents} with
$\Delta=1$. 
The scaling in Fig. \ref{S_LJ} supports the validity of our theory
for the elastic contact force given by Eq. \eqref{general:force}.

\subsection{$\Delta$-dependence of the critical exponents}
\label{delta:subsec}

\begin{figure}
\begin{center}
\includegraphics[height=14em]{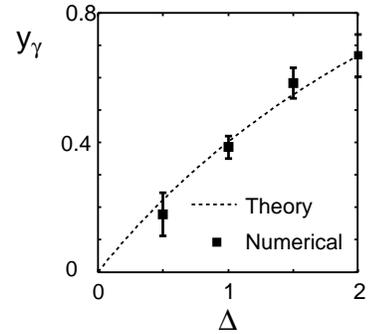}
\caption{ 
 Plot of  $y_\gamma$  versus $\Delta$ for the {\it polydisperse} system 
 ($N=8000$)
 with $D=2$.
 }
\label{yg:fig}
\end{center}
\end{figure}

\begin{figure}
\begin{center}
\includegraphics[height=14em]{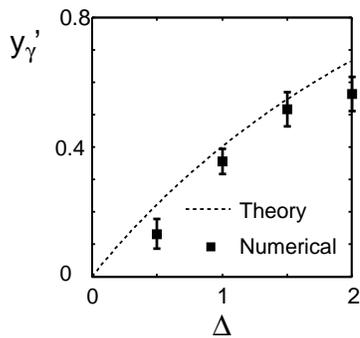}
\caption{ Plot of $y_\gamma'$ versus $\Delta$ for the {\it polydisperse} system 
 ($N=8000$)
with $D=2$. }
\label{ygd:fig}
\end{center}
\end{figure}

In order to verify  the $\Delta$-dependence of the critical exponents,
we plot $y_\gamma$ and $y_\gamma '$ versus $\Delta$
for the {\it polydisperse} system with $D=2$, $N=4000$, 
and the viscous force given by Eq. \eqref{dis:lin}
in Figs. \ref{yg:fig} and \ref{ygd:fig}, respectively.
Here, in order to obtain Figs. \ref{yg:fig} and \ref{ygd:fig},
 we have estimated $y_\gamma$ and $y_\gamma '$, respectively, 
from the shear stress $S(\dot \gamma,\phi)$ 
and the pressure $P(\dot \gamma,\phi)$ at $\phi=\phi_J$ as
\begin{eqnarray}
y_\gamma & = & 
\frac{\log (S(\dot \gamma_1,\phi_J)) - \log(S(\dot \gamma_2,\phi_J))}
{\log \dot \gamma_1 - \log \dot \gamma_2}, \label{eq20} \\
y_\gamma' & = & 
\frac{\log (P(\dot \gamma_1,\phi_J)) - \log(P(\dot \gamma_2,\phi_J))}
{\log \dot \gamma_1 - \log \dot \gamma_2},
\label{eq21}
\end{eqnarray}
for $(\dot \gamma_1, \dot \gamma_2)= 
  (5 \times 10^{-7}, 1.5 \times 10^{-6}),
  (1.5 \times 10^{-6}, 5 \times 10^{-6} ),
  (5 \times 10^{-6}, 1.5 \times 10^{-5} ),
  (1.5 \times 10^{-5}, 5 \times 10^{-5} ),
  (5 \times 10^{-7}, 5 \times 10^{-6}),
  (5 \times 10^{-6}, 5 \times 10^{-5})$, 
and plotted the averages of $y_\gamma$ and $y_\gamma '$
for different $(\dot \gamma_1, \dot \gamma_2)$ 
with the error bars, 
whose lengths are twice of the standard deviation
of $y_\gamma$ and $y_\gamma '$.
The exponent $y_{\gamma}$ reasonably agrees with the prediction
Eq. \eqref{exponents} in the wide range of $\Delta$. Although $y_{\gamma}'$ is a
little deviated from the theoretical prediction, we believe that the
deviation becomes smaller if we use the data for smaller $\dot\gamma$
and larger $N$.

\begin{figure}
\begin{center}
\includegraphics[height=12em]{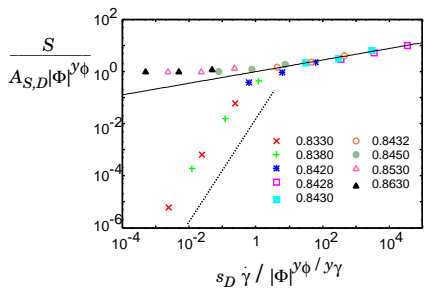}
\caption{ (Color online)
 Collapsed data of the shear rate dependence of the shear stress $S$
 for the {\it polydisperse} system ($N=8000$)
with $\Delta=0.5$ using the scaling law, Eq. (\ref{S:scale}), for $D=2$.
The dotted line and the solid line are proportional to 
$\dot{\gamma}^2$ and $\dot{\gamma}^{y_\gamma}$, respectively.
 The legends show the volume fraction $\phi$.
 The critical exponents are estimated from Eq. \eqref{exponents}
 as $y_\phi=1/2$ and $y_\gamma = 2/9$.
  }
\label{S_Dhalf}
\end{center}
\end{figure}

\begin{figure}
\begin{center}
\includegraphics[height=12em]{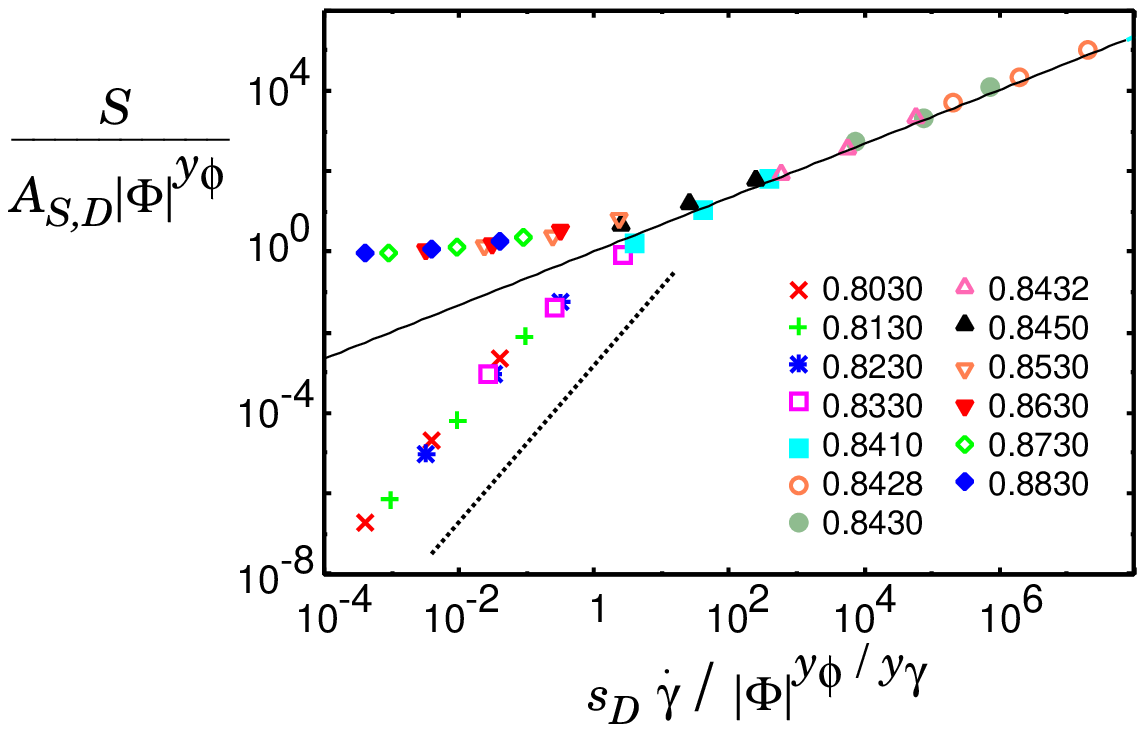}
\caption{ (Color online)
 Collapsed data of the shear rate dependence of the shear stress $S$
 for the {\it polydisperse} system ($N=8000$)
with $\Delta=2.0$ using the scaling law, Eq. (\ref{S:scale}), for $D=2$.
The dotted line and the solid line are proportional to 
$\dot{\gamma}^2$ and $\dot{\gamma}^{y_\gamma}$, respectively.
 The legends show the volume fraction $\phi$.
 The critical exponents are estimated from Eq. \eqref{exponents}
 as $y_\phi=2$ and $y_\gamma = 2/3$.
  }
\label{S_Dtwo}
\end{center}
\end{figure}

Figures \ref{S_Dhalf} and \ref{S_Dtwo} demonstrate whether 
the exponents  for $S$ predicted by Eq. \eqref{exponents} 
can be used for the scaling plots of 
$\Delta=0.5$ and $2$, respectively.
The shear rate $\dot{\gamma}$ is ranged between
$5\times 10^{-7}$ and $5 \times 10^{-5}$.
The amplitude and the adjustable parameter are
$(s_D,A_{s,D}) = 
(0.015, 0.05)$ for  $\Delta=0.5$
and $(0.05, 0.02)$ for  $\Delta=2.0$, respectively.
These scalings in Figs. \ref{S_Dhalf} and \ref{S_Dtwo} as well 
as the evaluated exponents by Eqs. (\ref{eq20}) and (\ref{eq21}) 
strongly support the validity of  the theoretical predictions in Eq. (\ref{exponents}) for 
arbitrary $\Delta$. Thus, our theory can be used for any $\Delta$.

\subsection{$\Phi$-dependence of critical variables}
\label{density:subsec}

\begin{figure}
\begin{center}
\includegraphics[height=14em]{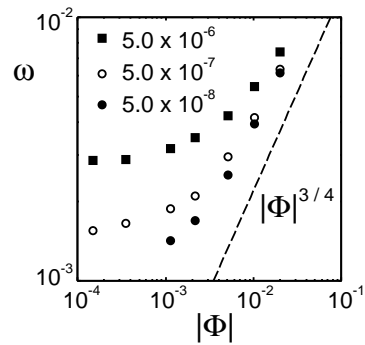}
\caption{ 
Plots of
 $\omega$ versus $\Phi$ of the {\it polydisperse} systems for
 $D=2$ and $\Delta = 3/2$
with $\dot{\gamma}=5\times 10^{-6}, 5\times 10^{-7}$, and $5\times 10^{-8}$.
  }
\label{omega:fig}
\end{center}
\end{figure}

Here, we examine $\Phi$-dependences of quantities predicted in Eqs. 
 (\ref{unjammed:scaling})  and (\ref{jammed:scaling})
 with Eq. \eqref{exponents}.
First, we plot $\omega$ versus $|\Phi|$
in the jammed phase for the two-dimensional {\it polydisperse} systems
($N=2000$) with $\Delta=3/2$ in Fig. \ref{omega:fig}.
We adopt Eq. (\ref{dis:lin}) for the viscous contact force.
Note that the corresponding results for $\Delta=1$ have been reported in  \cite{Otsuki08}.
As the shear rate $\dot \gamma$ decreases, $\omega$ approaches
$\omega \sim |\Phi|^{3/4}$ as predicted in 
Eq. (\ref{jammed:scaling})
with Eq. (\ref{exponents}).
There is a plateau in the small $\dot \gamma$ region,
but the value
of it decreases as $\dot\gamma^{z_\gamma}$ in the limit  $\dot \gamma \rightarrow 0$,
which can be predicted from Eq. (\ref{w:scale}) because
${\cal W}_+(x) \sim x^{z_\gamma}$ with $x=\dot \gamma / |\Phi|^{z_\phi / z_\gamma}
\rightarrow \infty$.

\begin{figure}
\begin{center}
\includegraphics[height=14em]{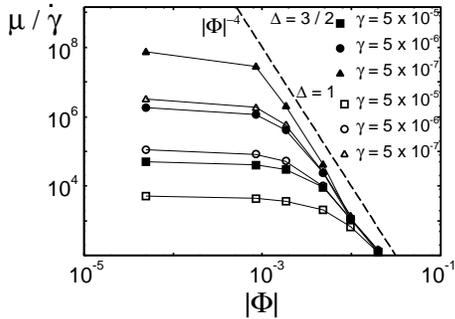}
\caption{ Plots of
$\mu/ \dot{\gamma}$ versus $\Phi$ of the two-dimensional 
{\it polydisperse} systems  for and $\Delta = 1$
and $3/2$ 
with $\dot{\gamma}=5\times 10^{-5}, 5\times 10^{-6}$, and $5\times 10^{-7}$.
 }
\label{eta_all}
\end{center}
\end{figure}

Next, we show  $\Phi$-dependence of the viscosity $\mu \equiv S / \dot\gamma$
in the unjammed phase, which is predicted as
$\mu \sim \dot \gamma |\Phi|^{-4}\sim |\Phi|^{-4}$
from
Eqs. (\ref{unjammed:scaling}) and (\ref{exponents}).
We note that the critical exponent for  $\mu$
is independent of $\Delta$ and $D$.
Figure \ref{eta_all} includes the data of $\mu / \dot \gamma$ as a function
of $|\Phi|$ for the {\it polydisperse} system ($N=2000$) 
with $\Delta =1$ and $3/2$.
We respectively adopt Eqs. \eqref{dis:lin} and \eqref{dis:her} for the viscous contact forces in 
in the cases of $\Delta=1$ and $3/2$.
Both of the data for $\Delta =1$ and $3/2$ satisfy the theoretical prediction 
$\mu/\dot \gamma \sim |\Phi|^{-4}$ as predicted,
although there is a plateau when $|\Phi| \rightarrow 0$.
The value of the plateau proportional to $\dot \gamma^{y_\gamma-2}$
can be understood by Eq.  (\ref{S:scale}) \cite{Otsuki08}.

There are some previous studies on the
divergence of the viscosity $\mu$.
For example, Losert et al. \cite{Losert}
observed exponents larger than 1 from their experiment.
The critical exponent of the shear viscosity 
for foams and emulsions in Ref. \cite{Olsson} is between 1 and 2.
For colloidal suspensions,
the viscosity is believed to diverge as $|\Phi|^{-2}$ \cite{Russel}. 
Garcia-Rojo et al. \cite{Garcia} reported that 
the scaled viscosity diverges as
$\mu / \sqrt{T} \sim (\phi_\mu-\phi)^{-1}$ for $\phi_\mu < \phi_J$
in two-dimensional elastic monodispersed hard-disks.
This result is contrast to our prediction
that the viscosity scaled by the temperature $\sqrt{T}$
diverges at point J as 
$\mu / \sqrt{T} \sim (\phi_J-\phi)^{-3}$,
obtained from Eq. \eqref{unjammed:scaling} with Eq. \eqref{exponents}.

\begin{figure}
\begin{center}
\includegraphics[height=14em]{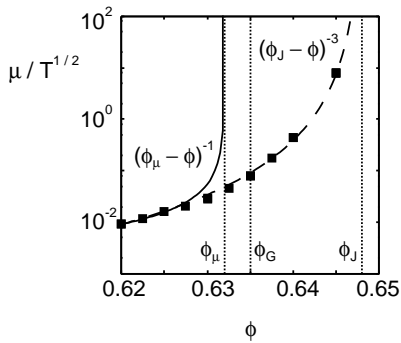}
\caption{ 
$\mu/ \sqrt{T}$ as a function of
$\phi$ for the three-dimensional {\it bidisperse} system ($N=4000$)
with $\Delta=1$ and $\dot \gamma = 5 \times 10^{-7}$,
 where the solid line and the dashed line are proportional to 
 $(\phi_\mu-\phi)^{-1}$
 and $(\phi_J-\phi)^{-3}$, respectively.
 }
\label{eta_L}
\end{center}
\end{figure}

\begin{figure}
\begin{center}
\includegraphics[height=14em]{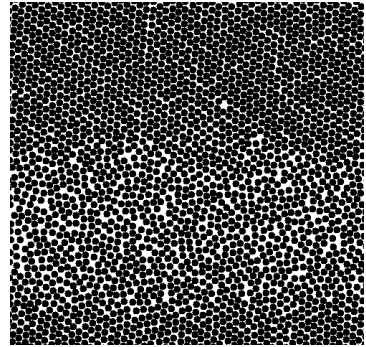}
\caption{ 
  The snapshot of the two-dimensional {\it monodisperse} system ($N=2401$)
  with $\Delta = 1$, $k=1.0$, and $\eta=0.00225$.
We use Eq. \eqref{dis:lin} for the viscous contact force. 
The shear rate is $\dot \gamma = 5 \times 10^{-5}$, 
and the density is $\phi=0.80$.
 }
\label{conf}
\end{center}
\end{figure}

In order to clarify whether our prediction is valid for sheared frictionless granular materials in the vicinity of the jamming transition,
we examine the possibility that the viscosity satisfies $\mu / \sqrt{T}
\sim (\phi_\mu-\phi)^{-1}$ with a fitting parameter $\phi_\mu$ \cite{Garcia} 
in Fig. \ref{eta_L} for
the {\it bidisperse} system ($N=4000$) with $D=3$, $\Delta=1$ and
$\dot \gamma = 5 \times 10^{-7}$.
Actually we can fit the data of our three-dimensional simulation 
by $\mu / \sqrt{T} \sim (\phi_\mu-\phi)^{-1}$, 
 but the viscosity is still finite even for $\phi>\phi_\mu=0.632$.
We, thus, conclude that the viscosity $\mu $ in the sheared granular materials
does not satisfy $\mu / \sqrt{T} \sim (\phi_\mu-\phi)^{-1}$,
and the behavior of $\mu $ is consistent with our prediction.
Here, we should note that
our theory  is no longer valid for the two-dimensional {\it monodisperse} system,
where the shear band occurs (Fig. \ref{conf}),
which is not assumed in our theory.

Berthier and Witten \cite{Berthier,Berthier09} reported that
the relaxation time in the zero-temperature limit
of the three-dimensional equilibrium {\it bidisperse} system
diverges at $\phi_G = 0.635$, which is smaller than $\phi_J = 0.639$.
Although this idea might be attractive to characterize universal feature of the jamming transition,  
we could not find such divergence, as shown in Fig. \ref{eta_L}, for frictionless sheared granular materials near the jamming transition.

\subsection{Nearly elastic cases}
\label{elastic:subsec}

\begin{figure}
\begin{center}
\includegraphics[height=12em]{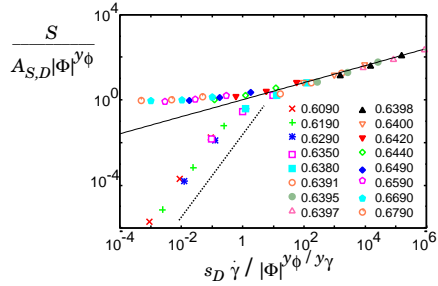}
\caption{ (Color online)
 Collapsed data of the shear rate dependence of the shear stress $S$
 in the nearly elastic {\it monodisperse} system ($N=2000$)
with $\Delta=1$ using the scaling law, Eq. \eqref{S:scale}, for $D=3$.
The the dotted line and the solid line are proportional to 
$\dot{\gamma}^2$ and $\dot{\gamma}^{y_\gamma}$, respectively.
 The legends show the volume fraction $\phi$.
 The critical exponents estimated from Eq. \eqref{exponents}
 are $y_\phi=1$ and $y_\gamma=2/5$.
  }
\label{S:elastic}
\end{center}
\end{figure}

One might think that our scaling theory is only valid  when the 
dissipation of the system is strong enough.
Indeed, we have used the viscous constant $\eta=1.0$,
which corresponds to the restitution coefficient $e=0.043$
for the linear spring model.
In order to check the validity of our theory in the nearly elastic system
($e \simeq 1$), we perform the numerical simulation of the three-dimensional
{\it monodisperse} system with $\Delta=1$ and $\eta=0.018$, which corresponds to
the restitution coefficient $e=0.96$.
We use the viscous contact force in Eq. \eqref{dis:lin} with the number of particles $N=2000$.
The shear rate $\dot \gamma$
is ranged between $5 \times 10^{-6}$ and $5 \times 10^{-4}$.
The amplitude and the adjustable parameter are given by
$(s_D,A_{s,D}) = 
(0.03, 0.04)$. 
The scaling plot of $S$ in this system is shown in Fig. \ref{S:elastic}.
This figure supports the validity of our scaling  even in the nearly elastic system.

\section{Pair correlation function}
\label{pair-correlation}


In this section, we discuss the behaviors of the spatial correlations in the vicinity of the jamming transition.
In particular, we focus on the critical behaviors at the first peak of the pair correlation function.
We restrict our interest to the {\it monodisperse} system where
each particle has an identical diameter $\sigma_0$.

\begin{figure}
\begin{center}
\includegraphics[height=14em]{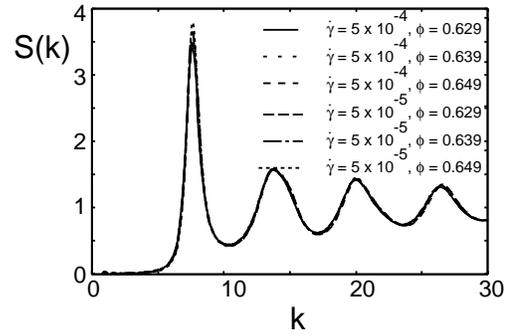}
\caption{ 
  Structure factor $S(k)$ in the three-dimensional {\it monodisperse} system 
  ($N=20000$)
  with $\phi = 0.629, 0.639, 0.649$
  and $\dot \gamma = 5 \times 10^{-4}$ and $5 \times 10^{-5}$. 
 }
\label{sk_all}
\end{center}
\end{figure}

\begin{figure}
\begin{center}
\includegraphics[height=14em]{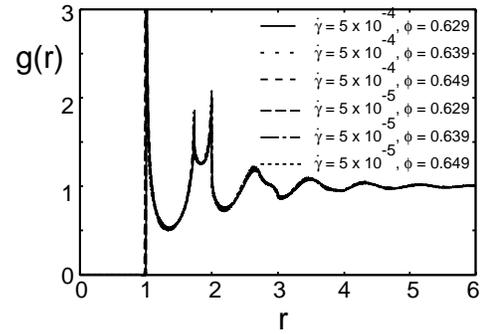}
\caption{ 
  Pair correlation function $g(r)$ in the three-dimensional {\it monodisperse} system ($N=20000$)
  with $\phi = 0.629, 0.639, 0.649$
  and $\dot \gamma = 5 \times 10^{-4}$ and $5 \times 10^{-5}$.
 }
\label{gr_all}
\end{center}
\end{figure}

\begin{figure}
\begin{center}
\includegraphics[height=14em]{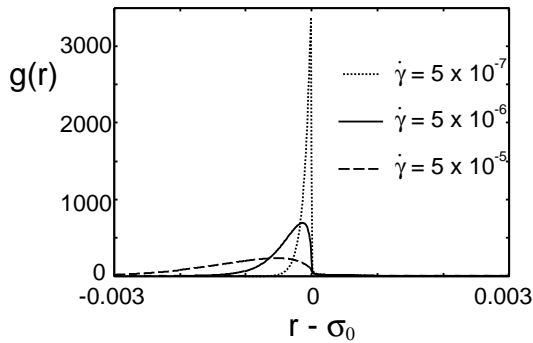}
\caption{ 
  The first peak of pair correlation function $g(r)$ 
  in the three-dimensional {\it monodisperse} system 
  ($N=20000$)
  with $\Delta=1$
  and $\phi = 0.639$
  for various shear rates $\dot \gamma$.
 }
\label{gr_ga}
\end{center}
\end{figure}

Let us discuss the spatial correlation of the density field. 
Figures \ref{sk_all} and \ref{gr_all} present the isotropic parts of 
the structure factor $S(k)$ and the pair correlation function $g(r)$
for the three-dimensional {\it monodisperse} system, respectively.
Here, $g(r)$ and $S(k)$ respectively satisfy \cite{Hansen}
\begin{eqnarray}
 g(r) & = &
 \frac{1}{ S_D r^{D-1} n}
\left < \frac{1}{N} \sum_i \sum_{j\neq i} \delta( r - r_{ij}) \right >,
\label{barg:def} \\
S(k) & = & 1 + (2 \pi)^{D/2} n \int_0^\infty dr \ r^{D-1} g(r) \frac{J_{D/2-1}(kr)}{(kr)^{D/2-1}},
\end{eqnarray}
where $J_{D/2-1}(kr)$ is the Bessel function,
$S_D$ is the surface area of the $D$-dimensional unit sphere
given by $S_D=2 \pi^{D/2}/ \Gamma (D/2)$ with the Gamma function $\Gamma (D/2)$.
We use $\Delta=1$, the viscous contact force given by Eq. \eqref{dis:lin},
and the number of the particles $N=20000$.
Because the isotropic parts of the spatial correlations
are dominant in our system, we only  focus on the critical properties 
of the isotropic parts in this section.
The first peak of $g(r)$ 
is  larger than 3 which is the maximum value of 
the vertical axis of Fig. \ref{gr_all}.
From Figs. \ref{sk_all} and \ref{gr_all}, it seems that there is no obvious
dependence of $S(k)$ and $g(r)$ on $\dot \gamma$ and $\phi$,
but the height
of the first peak $g_0$ of $g(r)$ strongly depends on $\dot\gamma$ and $\phi$.
We plot the dependence of the first peak of $g(r)$
on 
$\dot \gamma$ for small $r-\sigma_0$ region in 
Fig. \ref{gr_ga},
in which 
the first peak $g_0$ becomes higher and the width of the half-height of the
first peak $h_0$ becomes narrower as $\dot \gamma$ decreases.

The dependence of the peak on $\phi$ and $\dot \gamma$
can be roughly estimated
from our scaling law for the pressure $P$ in Eq. (\ref{P:scale}). 
Indeed,
the coordination number $Z$ and the pressure $P$, respectively, satisfy
\begin{eqnarray}
Z & = & \frac{ S_D n}{2} \int_0^{\sigma_0}dr r^{D-1} g(r), \label{Z:eq}\\
P & \simeq & 
\frac{ S_D n^2 k}{2}\int_0^{\sigma_0} dr r^D  (\sigma_0 - r)^\Delta
g(r). \label{P:eq}
\end{eqnarray} 
The derivation of these equations is briefly explained in Appendix B.
Since the first peak near $\sigma_0$ is characterized by the peak value $g_0$
and the width $h_0$, Eqs. \eqref{Z:eq} and \eqref{P:eq} are approximated by
\begin{eqnarray}
Z & \sim & S_D n \int_{\sigma_0-h_0}^{\sigma_0} r^{D-1}g_0 
\nonumber \\
& \sim &  g_0 h_0 \{ 1 + O(h_0)\}, 
\label{Z:app} \\
P & \sim & 
S_D n^2 \int_{\sigma_0 - h_0} ^{\sigma_0} dr r^D  
 k (\sigma_0 - r)^\Delta g_0, \nonumber \\
& \sim &
  h_0^{\Delta+1} g_0 \{ 1 + O(h_0) \}.
 \label{P:app}
\end{eqnarray} 
With the aid of  
Eq. \eqref{Z:app}, we find 
that $g_0$ and $h_0$ satisfy the relation near the point J 
\begin{eqnarray}
g_0 h_0 \sim const. ,
\label{g_h}
\end{eqnarray}
where  we have used the known result on the coordination number 
$Z \simeq 2D$ for the frictionless granular particles near the point J. 
Substituting Eq. \eqref{g_h} into Eq. \eqref{P:app}, we obtain the 
relation between $g_0$ and the pressure $P$ 
\begin{eqnarray}\label{Eq28}
g_0 \sim P^{-1/\Delta}. \label{g0_P}
\end{eqnarray}
From Eqs. \eqref{unjammed:scaling}, \eqref{exponents} and \eqref{Eq28}, 
the height of the first peak value $g_0$ in the unjammed phase is 
given by
\begin{eqnarray}
g_0 \sim \dot \gamma ^{-2/\Delta} |\Phi|^{4/\Delta}. \label{g_0:unjammed}
\end{eqnarray}
Similarly, from Eqs. (10), \eqref{exponents} and (\ref{Eq28}) $g_0$ satisfies
\begin{eqnarray}
g_0 \sim |\Phi|^{-1} \label{g_0:jammed}
\end{eqnarray}
in the jammed phase,
which is consistent with the numerical result of the unsheared jammed system
\cite{OHern03}.
At the critical point, i.e. $\Phi=0$, thus, $g_0$ is given by
\begin{eqnarray}
g_0 \sim \dot \gamma^{-2/(\Delta+4)} \label{g_0:critical}
\end{eqnarray}
from Eqs. (11), \eqref{exponents}, and \eqref{Eq28}.

\begin{figure}
\begin{center}
\includegraphics[height=14em]{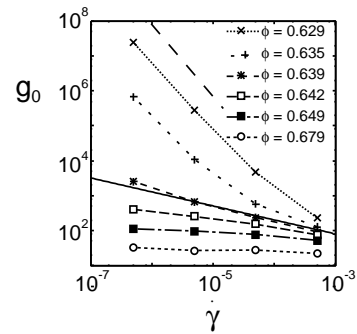}
\caption{ 
  The height of the first peak  $g_0$ of $g(r)$
  as a function of $\dot \gamma$
  for the three-dimensional {\it monodisperse} systems ($N=20000$)
    with $\Delta=1$
  for various densities $\phi$.
    The critical density is $\phi_J=0.639$.
    The solid line is proportional to $\dot \gamma^{-2/(\Delta+4)}$.
    The dashed line is proportional to $\dot \gamma^{-2/\Delta}$.
 }
\label{g0_ga}
\end{center}
\end{figure}

In order to check our predictions in Eqs. 
\eqref{g_0:unjammed}--\eqref{g_0:critical}, 
we plot $g_0$ versus $\dot \gamma$
for various densities in Fig. \ref{g0_ga}.
The numerical data are consistent with the theoretical prediction in which
$g_0$ is proportional to $\dot \gamma^{-2/\Delta}$ in the unjammed phase (see Eq.(\ref{g_0:unjammed})),
but $g_0$ is almost a constant in the jammed phase (see Eq.\eqref{g_0:jammed}), and
$g_0$ satisfies $g_0 \sim \dot \gamma^{-2/(\Delta+4)}$ at point J (see Eq.\eqref{g_0:critical}).

\begin{figure}
\begin{center}
\includegraphics[height=14em]{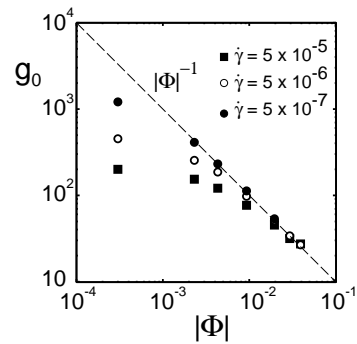}
\caption{ The plots of
  $g_0$ versus $|\Phi|$
  in the jammed phase for the three-dimensional {\it monodisperse} system 
($N=20000$)
  with $\Delta=1$
  for various shear rates $\dot \gamma$.
The dashed guide line is proportional to $|\Phi|^{-1}$.
 }
\label{g0_phi_jammed}
\end{center}
\end{figure}

\begin{figure}
\begin{center}
\includegraphics[height=14em]{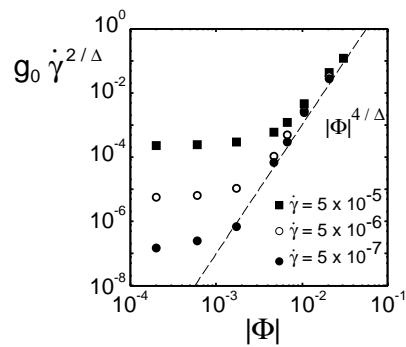}
\caption{ 
The plots of  $g_0 \dot \gamma^{2 / \Delta}$ versus $|\Phi|$
  in the unjammed phase for the {\it monodisperse} system 
  ($N=20000$) with $\Delta=1$
  for various shear rates $\dot \gamma$.
The dashed guide line is proportional to $|\Phi|^{4/\Delta}$.
 }
\label{g0_phi_unjammed}
\end{center}
\end{figure}

Figure  \ref{g0_phi_jammed}  examines the validity of 
Eq. \eqref{g_0:jammed} in the zero shear limit of the jammed phase,
in which $g_0$ seems to satisfy $g_0\sim 1/|\Phi|$.
On the other hand,  $g_0$ tends to satisfy $g_0 \dot \gamma^{2/\Delta}\sim |\Phi|^{\Delta/4}$
in the unjammed phase as in Eq. \eqref{g_0:unjammed} (see
 Fig. \ref{g0_phi_unjammed}).

\section{Discussion and Conclusion}

\label{Discussion:sec}

This section consists of two parts. In the first part, we will discuss our results, and we will conclude our work in the second part.
In the first part, let us compare our results with those by Hatano
\cite{Hatano08}, discuss the growing length scale near point J,
the long-range spatial correlation,
and the scaling laws for the Langevin dynamics
and the frictional particles.  

\subsection{Discussion}

\begin{figure}[htbp]
\begin{center}
\includegraphics[height=14em]{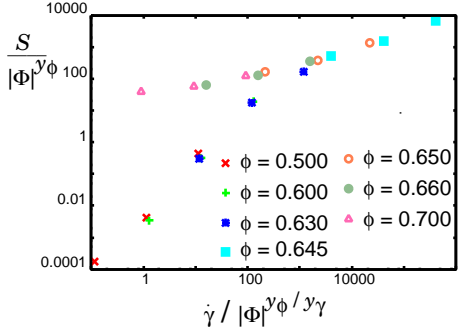}
\caption{ 
The scaling of the shear stress $S/|\Phi|^{y_\phi}$
for the {\it polydisperse} system 
  ($N=4000$) with $\Delta=1$
using Eq. (\ref{exponents_Hatano_L})
as a function of the scaled shear rate 
$\dot \gamma /|\Phi|^{y_\phi / y_\gamma}$ for
$10^{-4} \leq \dot \gamma \leq 10^{-1}$, $0.5 \leq \phi \leq 0.7$ with $D=3$.
}
\label{S_large}
\end{center}
\end{figure}
\begin{figure}[htbp]
\begin{center}
\includegraphics[height=14em]{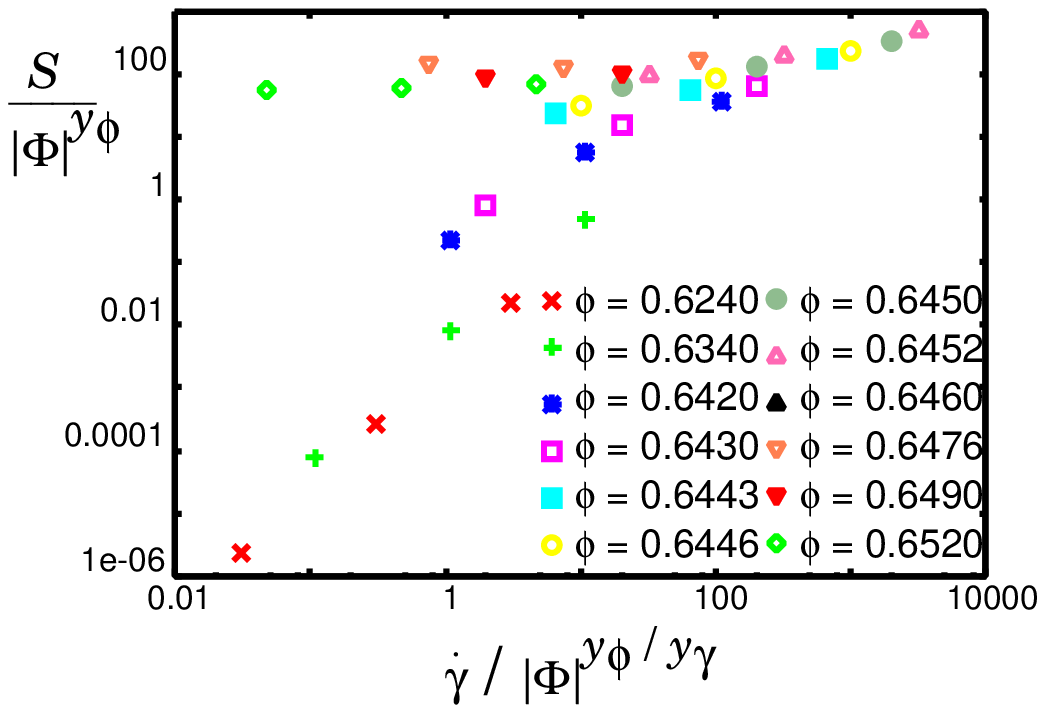}
\caption{ 
The scaling of the shear stress $S/|\Phi|^{y_\phi}$
for the {\it polydisperse} system
  ($N=4000$) with $\Delta=1$
using Eq. (\ref{exponents_Hatano_L})
as a function of the scaled shear rate 
$\dot \gamma /|\Phi|^{y_\phi / y_\gamma}$ for
 $ 5 \times 10^{-7} \leq \dot \gamma \leq 5 \times 10^{-5}$,
 $0.624 \leq \phi \leq 0.652$ with $D=3$.}
\label{S_small}
\end{center}
\end{figure}
\begin{figure}[htbp]
\begin{center}
\includegraphics[height=14em]{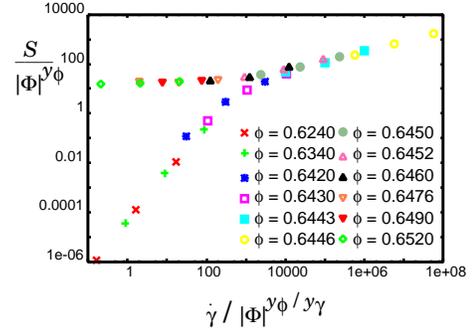}
\caption{ 
The scaling of the shear stress $S/|\Phi|^{y_\phi}$
for the {\it polydisperse} system
  ($N=4000$) with $\Delta=1$
using Eq. (\ref{exponents})
as a function of the scaled shear rate 
$\dot \gamma /|\Phi|^{y_\phi / y_\gamma}$ for
$ 5 \times 10^{-7} \leq \dot \gamma \leq 5 \times 10^{-5}$,
 $0.624 \leq \phi \leq 0.652$ with $D=3$.
}
\label{S_small_O}
\end{center}
\end{figure}

Let us compare our results with those by Hatano \cite{Hatano08}.
Hatano estimated the values of the exponents
as
\begin{eqnarray}
x_{\Phi}&=&2.5, \quad x_{\gamma}=1.3, \quad y_{\Phi}=1.2, \nonumber \\
y_{\gamma} &=&0.63, 
 \quad y_{\Phi}' = 1.2, \quad y_{\gamma}'=0.57,
\label{exponents_Hatano_L}
\end{eqnarray} 
for $\Delta=1$, and 
\begin{eqnarray}
x_{\Phi}&=&3.2, \quad x_{\gamma}=1.3, \quad y_{\Phi}=1.8, 
\nonumber \\
y_{\gamma}&=&0.75,
 \quad y_{\Phi}' = 1.8, \quad y_{\gamma}'=0.72,
\label{exponents_Hatano_H}
\end{eqnarray} 
for $\Delta=3/2$.
The system analyzed in Ref. \cite{Hatano08}
corresponds to our {\it polydisperse} system, but
these values are different from those 
in Eq. (\ref{exponents}).
Here, let us clarify the origin of the differences.
(i) The estimation of the critical exponents strongly depends
on  the range of $\dot \gamma$.
The value of $\phi_J$ and the range of $\dot \gamma$
in Ref. \cite{Hatano08} are larger than ours. 
If we adopt $\phi_J$ and the range of $\dot \gamma$
in Ref. \cite{Hatano08},
we can recover his scaling laws 
in Eqs. (\ref{exponents_Hatano_L}) and (\ref{exponents_Hatano_H})
as shown in Fig. \ref{S_large},
while we find the obvious violation of his scaling (Fig. \ref{S_small})  in the small $\dot \gamma$ region ($\dot \gamma < 10^{-4}$).
On the other hand, our scaling is still valid as shown in Fig. \ref{S_small_O} for $\dot\gamma<10^{-4}$.
In order to extract the critical properties, 
it is needless to say that we should use the data in the small $\dot \gamma$ region.
Hence, our prediction for the exponents
is more appropriate than Hatano's estimation.
(ii) 
The estimated exponents 
in Eqs. (\ref{exponents_Hatano_L}) and (\ref{exponents_Hatano_H})
cannot be valid even 
when we adjust the value of $\phi_J$ as the fitting parameter 
in the small $\dot \gamma$ region.
The value of $\phi_J$ 
is well established
by the simulation of the sphere packing
for the {\it bidisperse} system and the {\it monodisperse} system
  \cite{OHern02,OHern03}.
The scaling plots using the estimated value of $\phi_J$
(Figs. \ref{S_L_2dis} and \ref{S_mono})
evidently support the validity of our prediction.

There are some studies to indicate the diverging time scale
near point J \cite{Wyart,Hatano08.3}.
Indeed, in the scaling relations of $T, S, P$, and $\omega$ 
in Eqs.  (\ref{T:scale})--(\ref{w:scale})
with Eq. (\ref{exponents}),
the shear rate $\dot \gamma$ is scaled by the same diverging time
scale $\tau \sim |\Phi|^{-(\Delta +4)/2}$.
For conventional critical phenomena,
the divergence of the time scale is associated with
the diverging length scale. Therefore,
there are many papers to discuss growing   length scale 
in the vicinity of the jamming transition
\cite{Abate06, Abate07, Dauchot08, Watanabe, Olsson, Hatano08.2, Hatano08.3, Wyart,
Head}.
However, the length scale in the previous studies 
is as much as the size of several particles in their papers,
and it is not clear whether the length scale diverges at point J. 
We, thus, conjecture that the length scale might be unrelated to determine
the critical exponents.
The success of our mean field theory supports the validity of this conjecture.


In relatively dilute sheared systems, we know 
the existence of the long-range
correlation \cite{Lutsko85, Lutsko02, Wada, Otsuki09}.
The existence  of the long-range correlation is only confirmed  
in the relatively dilute systems such as $\phi\le 0.50$ for the three dimensional case.
We, thus, still do not know whether there is
the long-range correlation in the jammed systems, and the role of the correlation.
We will discuss the long-range correlation in the jammed systems elsewhere.


The similar scaling relations with different values of the exponents 
are observed in the zero temperature limit of Langevin thermostat system,
where the Newtonian law $S \propto \dot \gamma$ is held 
in the unjammed region \cite{Olsson}.
However, we cannot extend our simple theory to this system
because the characteristic frequency defined by $\omega=\dot \gamma S/(nT)$
is always constant in this system, 
which differs from the scaling relation
 (\ref{w:scale}).
 We will discuss the results of this situation elsewhere.


In this paper, we restrict our interest to the frictionless particles.
When the particles have friction,
the situation will be changed completely.
For examples, 
the critical density $\phi_J$ of the frictional particles 
in the static granular packings,
becomes smaller than
that of the frictionless particles \cite{Hecke}, 
and depends on the packing process \cite{Makse, Inagaki}.
Thus, critical properties of the scaling 
in the sheared dynamical systems of frictional particles 
should differ from the frictionless assemblies.
This also will be our future work.

\subsection{Conclusion}

In conclusion, 
we have extensively checked the validity of the mean field theory proposed 
in Ref. \cite{Otsuki08} numerically, and we demonstrate that most of all our
numerical results are consistent with the theoretical predictions 
in Eq. \eqref{exponents}.
Thus, we may conclude that the jamming transition for frictionless sheared granular materials is a continuous transition
in which the critical exponents are independent of the spatial dimension and are determined by the local elastic force between contacted grains.
We also confirm that the viscosity diverges at point J as $(\phi_J-\phi)^{-4}$.
Essential new findings beyond 
Ref. \cite{Otsuki08} are the critical behaviors of the first peak of the pair correlation function given by 
Eqs. \eqref{g_0:unjammed}, \eqref{g_0:jammed} and \eqref{g_0:critical}, which are also consistent with our numerical simulation.

\begin{acknowledgments}
We thank S. Luding, S. Sasa, B. Kim, and S.-H. Chong 
for the valuable discussion.
This work is partially supported by Ministry of Education, Culture, Science and Technology (MEXT), Japan
 (Grant Nos. 21015016, 21540384 and 21540388) and the Grant-in-Aid for the global COE program
"The Next Generation of Physics, Spun from Universality and Emergence"
from the Ministry of Education, Culture, Sports, Science and
Technology (MEXT) of Japan.
The numerical calculations were carried out on Altix3700 BX2 at YITP in Kyoto University.
\end{acknowledgments}

\appendix

\section{Derivation of the values of the critical exponents}
\label{exponents:app}

In this appendix, we theoretically determine the values of
the critical exponents in Eq (\ref{exponents}). The derivation
is parallel to that in Ref. \cite{Otsuki08}, but contains some generalizations with the help of a simplified argument.

At first, we should note that the scaling functions ${\cal T}_{\pm}(x)$, ${\cal S}_{\pm}(x)$, 
 ${\cal P}_{\pm}(x)$, and ${\cal W}_{\pm}(x)$ satisfy
\begin{eqnarray}\label{jammed}
\lim_{x\rightarrow 0}{\cal T}_+(x) & =& x, \quad 
\lim_{x\rightarrow 0}{\cal S}_{+}(x) = 1, \nonumber \\
\lim_{x\rightarrow 0}{\cal P}_{+}(x) & = &1, \quad 
\lim_{x\rightarrow 0}{\cal W}_{+}(x) = 1,
\end{eqnarray}
\begin{eqnarray}\label{unjammed}
\lim_{x\rightarrow 0}{\cal T}_{-}(x) & = & x^2, {\quad}
\lim_{x\rightarrow 0}{\cal S}_{-}(x) = x^2, \nonumber \\
\lim_{x\rightarrow 0}{\cal P}_{-}(x) & = & x^2, {\quad}
\lim_{x\rightarrow 0}{\cal W}_{-}(x) = x,
\end{eqnarray}
\begin{eqnarray}\label{critical}
\lim_{x\rightarrow \infty}{\cal T}_{\pm}(x) & = & x^{x_\gamma}, \
\lim_{x\rightarrow \infty}{\cal S}_{\pm}(x) = x^{y_\gamma}, \nonumber \\
\lim_{x\rightarrow \infty}{\cal P}_{\pm}(x) & = & x^{y_\gamma'}, \
\lim_{x\rightarrow \infty}{\cal W}_{\pm}(x) = x^{z_\gamma}.
\end{eqnarray}
The scaling relations 
\eqref{unjammed:scaling}--\eqref{critical:scaling} are
obtained from Eqs. \eqref{jammed}--\eqref{critical}
with Eqs. \eqref{T:scale}--\eqref{w:scale}.

Let us assume that the inverse of the shear rate $\dot\gamma^{-1}$
can be scaled by a characteristic time scale.
 Therefore we can assume that
the ratios between the exponents $x_\phi / x_\gamma$, $y_\phi / y_\gamma$, $y_\phi' / y_\gamma'$
and $z_\phi / z_\gamma$ in the scaling laws Eqs. \eqref{T:scale}--\eqref{w:scale}
are common as
\begin{equation}\label{alpha}
\alpha = \frac{x_\phi}{x_\gamma} = \frac{y_\phi}{y_\gamma} = \frac{y_\phi'}{y_\gamma'} = \frac{z_\phi}{z_\gamma}.
\end{equation}
In other words, the characteristic time scale $\tau$
exhibits critical slowing down as $\tau \sim |\Phi|^{-\alpha}$.
This property has already been indicated by Ref. \cite{Hatano08}.
It should be noted that 
we can determine the exponents without this assumption \cite{Otsuki08}.

Substituting Eq.  (\ref{unjammed:scaling})
into Eq. (\ref{omega}) with Eq. \eqref{alpha}, 
we obtain the relation between the exponents as
\begin{equation} \label{A5}
z_\phi = y_\phi - x_\phi + \alpha.
\end{equation}

Let us assume that the pressure in the jammed phase 
($\Phi>0$ and $\dot \gamma \rightarrow 0$)  converges to
that of unsheared jammed phase satisfying $P \sim \Phi ^\Delta$ \cite{OHern03}.
Hence, comparing the scaling property of $P$ in Eq. (\ref{jammed:scaling})
with $P \sim \Phi ^\Delta$, we find the relation
\begin{equation}
y_\phi' = \Delta.
\end{equation}

We also assume that Coulomb's frictional law is held
in granular systems \cite{Hatano07.2}. Thus,  $S/P$ is independent of $\Phi$
\cite{Otsuki08}
and we obtain
\begin{equation}
y_\phi = y_\phi',
\end{equation}
with the aid of Eq. \eqref{jammed:scaling}.

We can use the properties of  the cut off frequency $f_c$ in 
the density of state in the jammed phase, which satisfies $f_c \sim \sqrt{P}$
\cite{Wyart}. 
Since we expect that there is only one time scale, it is reasonable to assume that the characteristic frequency $\omega$ in the limit
$\dot{\gamma} \rightarrow 0$ can be scaled 
by the cutoff frequency $f_c$.
Thus, we obtain
\begin{equation}
z_\phi = \frac{1}{2} y_\phi'.
\end{equation}

Finally, let us use a similar argument on the characteristic frequency $\omega$ in the unjammed phase to that in the jammed phase.
 Since the characteristic frequency $\omega$ 
 is estimated as $\omega \sim 
\sqrt{T/m}/l(\Phi)$ with the mean free path $l(\Phi)$,
which is evaluated as 
$(\sigma /D\phi_J) |\Phi|$ in the vicinity of point J,
 we obtain
\begin{equation} \label{A9}
x_\phi = 2 z_\phi + 2,
\end{equation}
where we have used Eq. (\ref{unjammed:scaling}) .

From Eqs. \eqref{A5}--\eqref{A9}, the exponents $\alpha$, $x_\phi$, $y_\phi$, $y_\phi'$, 
and $z_\phi$ are given by
\begin{eqnarray} \label{exponents:phi}
\alpha & = & \frac{\Delta + 4}{2}, \quad x_{\Phi}=2+\Delta, \quad y_{\Phi}=\Delta, 
\nonumber \\
\quad y_{\Phi}' & = & \Delta, \quad
z_{\Phi}=\frac{\Delta}{2}.
\end{eqnarray} 
The exponents $x_\gamma$, $y_\gamma$, $y_\gamma'$, 
and $z_\gamma$ are obtained from  Eqs. (\ref{alpha}) and (\ref{exponents:phi}).
Hence, we have determined all the critical exponents as Eq. (\ref{exponents}).

\section{The expressions of $P$ and $Z$ by $g(r)$}
\label{expresion:P_Z}

In this appendix, we derive Eqs. \eqref{Z:eq} and \eqref{P:eq}.
The coordination number $Z$ is given by $Z = M/N$, 
where $M$ is the number of the points of contact.
Since $M$ is given by the number of the pairs 
whose distances are smaller than $\sigma_0$, $Z$ is given by
\begin{eqnarray}
Z & = & \frac{1}{N} \int_0^{\sigma_0} dr
\left < \frac{1}{2}\sum_i \sum_{j\neq i} \delta( r - r_{ij}) \right >.
\label{Z:eq2} 
\end{eqnarray} 
Substituting Eq. \eqref{barg:def} into this equation,
we obtain Eq. \eqref{Z:eq}.

Since the contribution to the pressure $P$ from the elastic force is 
dominant in our system,
we approximate $P$ in Eq. \eqref{P:ex}
with the aid of  Eq. \eqref{barg:def} as
\begin{eqnarray}
P & \simeq & 
\frac{1}{2DV} \left < \sum_i \sum_{j \neq i} r_{ij} 
 f_{\rm el}(r_{ij}) 
 \right > \nonumber \\
 & = &
\frac{1}{2DV} \int_0^{\infty} dr r  f_{\rm el}(r)
\left < \sum_i \sum_{j \neq i} r_{ij} 
\delta \left( r - r_{ij} \right )
 \right >  \nonumber \\
& = &
\frac{ S_D n^2 }{2}\int_0^{\infty} dr r^D  f_{\rm el}(r)
g(r).
\label{P:eq2}
\end{eqnarray} 
Substituting Eq. \eqref{elastic:force} into Eq. \eqref{P:eq2},
we obtain Eq. \eqref{P:eq}.

\end{document}